\documentclass[manuscript]{acmart}
\AtBeginDocument{%
  \providecommand\BibTeX{{%
    \normalfont B\kern-0.5em{\scshape i\kern-0.25em b}\kern-0.8em\TeX}}}

\setcopyright{acmcopyright}
\copyrightyear{2023}
\acmYear{2023}
\acmDOI{XXXXXXX.XXXXXXX}

\acmConference[arXiv]{arXiv}{January 2023}{Ithaca, New York}
\acmBooktitle{arXiv, January 2023, Ithaca, New York}
\acmPrice{15.00}
\acmISBN{978-1-4503-XXXX-X/18/06}

\usepackage{hhline}

\usepackage{makecell}

\usepackage{multirow}
\usepackage{multicol}

\usepackage{tablefootnote}

\usepackage{longtable}

\usepackage{ragged2e}

\usepackage{array}

\usepackage{siunitx}

\usepackage{amsmath}

\usepackage{enumitem}
\setlist[itemize]{noitemsep,partopsep=0pt,topsep=0pt,parsep=0pt}
\setlist[enumerate]{noitemsep,partopsep=0pt,topsep=0pt,parsep=0pt}

\begin{document}

\title{What Do Children and Parents Want and Perceive in Conversational Agents? Towards Transparent, Trustworthy, Democratized Agents}

\author{Jessica Van Brummelen}
\email{jess@csail.mit.edu}
\orcid{0000-0002-4831-6296}
\affiliation{%
  \institution{Massachusetts Institute of Technology}
  \streetaddress{77 Massachusetts Ave}
  \city{Cambridge}
  \state{Massachusetts}
  \country{USA}
  \postcode{02139}
}

\author{Maura Kelleher}
\email{maurakel@mit.edu}
\affiliation{%
  \institution{Massachusetts Institute of Technology}
  \streetaddress{77 Massachusetts Ave}
  \city{Cambridge}
  \state{Massachusetts}
  \country{USA}
  \postcode{02139}
}

\author{Mingyan Claire Tian}
\email{mt1@wellesley.edu }
\affiliation{%
  \institution{Wellesley College}
  \streetaddress{106 Central St}
  \city{Wellesley}
  \state{Massachusetts}
  \country{USA}
  \postcode{02481}
}

\author{Nghi Hoang Nguyen}
\email{nghin@mit.edu}
\affiliation{%
  \institution{Massachusetts Institute of Technology}
  \streetaddress{77 Massachusetts Ave}
  \city{Cambridge}
  \state{Massachusetts}
  \country{USA}
  \postcode{02139}
}

\renewcommand{\shortauthors}{Van Brummelen, et al.}

\begin{abstract}
Historically, researchers have focused on analyzing WEIRD, adult perspectives on technology. This means we may not have technology developed appropriately for children and those from non-WEIRD countries. In this paper, we analyze children and parents from various countries’ perspectives on an emerging technology: conversational agents. We aim to better understand participants' trust of agents, partner models, and their ideas of ``ideal future agents'' such that researchers can better design for these users. Additionally, we empower children and parents to program their own agents through educational workshops, and present changes in perceptions as participants create and learn about agents. Results from the study (n=49) included how children felt agents were significantly more human-like, warm, and dependable than parents did, how participants trusted agents more than parents or friends for correct information, how children described their ideal agents as being more artificial than human-like than parents did, and how children tended to focus more on fun features, approachable/friendly features and addressing concerns through agent design than parents did, among other results. We also discuss potential agent design implications of the results, including how designers may be able to best foster appropriate levels of trust towards agents by focusing on designing agents' competence and predictability indicators, as well as increasing transparency in terms of agents' information sources.
\end{abstract}





\begin{teaserfigure}
  \includegraphics[width=\textwidth]{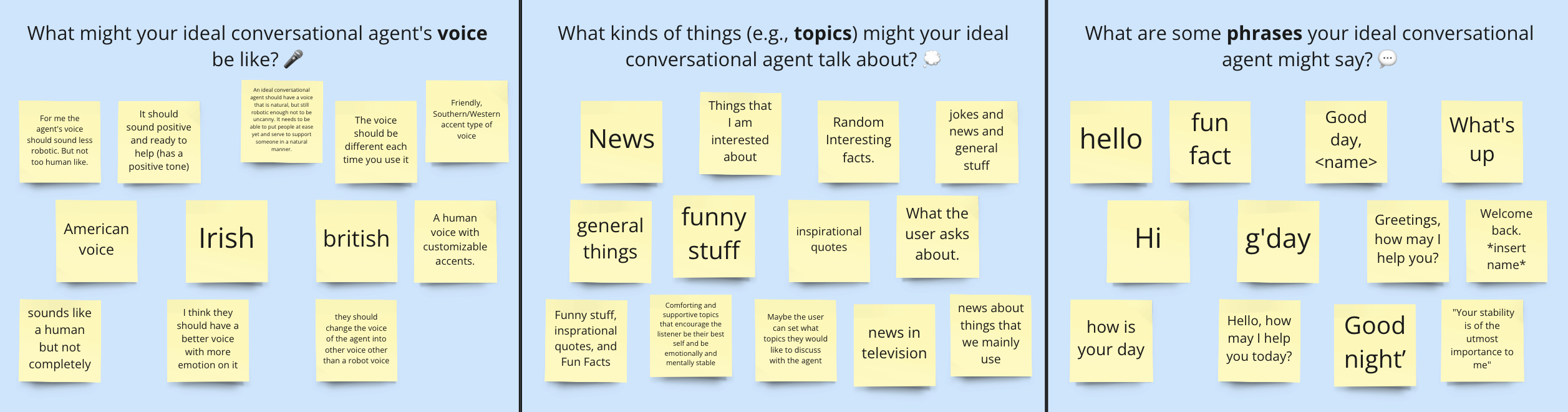}
  \caption{A portion of child participant responses during an ideation design session about their ideal conversational agents.}
  \Description{Virtual sticky-note responses under three headings. Under the heading, ``What might your ideal conversational agent's voice be like?'', responses included, ``For me the agent's voice should sound less robotic. But not too human like.'', ``American voice'', and ``A human voice with customizable accents.'', among others. Under the heading, ``What kinds of things (e.g., topics) might your ideal conversational agent talk about?'', responses included, ``Things that I am interested about'', ``funny stuff'' and ``news in television'', among others. Under the heading, ``What are some phrases your ideal conversational agent might say?'', responses included, ``Good day, <name>'', ``Hello, how may I help you today?'' and ``fun fact'', among others.}
  \label{fig:miro-board-teaser}
\end{teaserfigure}

\maketitle

\section{Introduction}

Conversational artificial intelligence (AI)---or the ability of a computer program to understand human language and respond accordingly---is ripe with potential. Imagine a conversational agent engaging children in learning history with a virtual Rosa Parks, or 
an agent providing constant, accurate healthcare answers to those in need. With recent major advances in natural language processing and automatic speech recognition %
these ideas are not far-fetched \cite{attention-transformers, gpt3, bert, nlp-universal-transfer-learning, asr-generative-raw-audio, asr-transformer, foundation-models}. %
Nonetheless, current agents, like Google Home, Apple's Siri and Amazon Alexa, still misrecognize speech and misunderstand intent \cite{siri-fail-ranker,alexa-fail-reddit,siri-fail-ranker}. %
For instance, researchers found speech recognition systems by Amazon, Google, IBM and Microsoft did substantially worse when recognizing black speakers versus white \cite{racial-biases-asr}. Others have found significant gender biases in embeddings %
\cite{gender-bias-elmo,gender-bias-embeddings}. Biases in AI systems are widespread, %
and if users are not aware of such flaws, %
there could be serious implications, including misinformation being spread, human bias being compounded, and users unwittingly acting on incorrect advice \cite{conv-ai-ethical-considerations}. %

Ideally, agents would be developed to portray the reality of their abilities and limitations to their users through effective %
design. In a study with AI decision-aids, researchers describe %
how if users are too averse to technology's advice and information, they cannot truly benefit from using the technology. However, if they are too appreciative, users may make ill-informed decisions when technology presents incorrect %
information \cite{algorithmic-aversion-clinical-deployment}. By portraying conversational agents in an honest way through design, discrepancies between users' expectations of agents---or their agent ``partner models''---and the reality of agents can be reduced, which can also reduce user frustration \cite{partner-models-doyle}.

In our study, we investigate users' perceptions of agents, including their partner models %
and trust. The results revealed how for certain aspects of agents---including warmth, human-likeness and dependability---children perceived agents differently than parents. Participants' general trust of agents' correctness (compared to other people's and systems' correctness), however, was similar for both children and parents. In general, people trusted agents more than their friends and parents. Based on these results and others, we discuss agent design recommendations to foster appropriate levels of trust of agents.%

Historically, human-computer interaction research has largely recruited participants from Western, Educated, Industrialized, Rich and Democratic (WEIRD) countries, who comprise less than 12\% of the world's population \cite{most-people-not-weird,how-weird-is-chi}. This means many of the design recommendations developers use are likely biased towards this population. Furthermore, a large portion of software developers reside in WEIRD countries \cite{acm-idc-number-of-developers,idc-number-of-developers,github-number-of-developers}, meaning technology development is likely further biased towards the WEIRD population. In order to address this, and develop technology meaningful and relevant to more of the world, %
researchers have developed different strategies. One strategy involves including more participants from non-WEIRD countries and developing recommendations based on wider demographics \cite{weird-hci-chi}. We utilize this strategy through involving participants from non-WEIRD and WEIRD countries, investigating their perceptions of agents, and asking them how they envision their ``ideal conversational agents''. The results and recommendations aim to provide agent designers with perspectives from those from different countries and generations. %

Another strategy to reduce the gap between non-WEIRD- and WEIRD-centric technology is to empower those from non-WEIRD countries to develop their own technology. %
There are a number of tools that help enable nearly anyone to develop technology, many of which utilize visual or block-based coding \cite{visual-block-based-coding}. These tools have largely been born out of the constructionist movement in education, which encourages the use of low-floor, high-ceiling programming tools to empower a wide variety of people to learn to program, including other underrepresented groups in the technology sector, like children \cite{papert-mindstorms}. Scratch, for instance, allows children to program their own web-based animations using block-based coding \cite{scratch}. Other low-floor platforms enable users to develop conversational agents, including the Flow Editor and Alexa Blueprints \cite{robot-flow-editor,alexa-blueprints}. The MIT App Inventor platform allows users to develop fully-fledged apps, which can be deployed to mobile devices' app stores \cite{democratizing-cs-app-inventor}, %
as well as conversational agents, which can be deployed to Amazon Alexa devices, through ``ConvoBlocks'' \cite{sm-summary-fall-symp,vanbrummelen-sm,convoblocks}. %

In this paper, we aim to democratize conversational agent technology to young learners from various countries and their parents through an educational intervention with the ConvoBlocks platform. This intervention empowers students to develop their own agents. We adopt ConvoBlocks in our study, as it is open-source and has a low barrier to creating deployable agents \cite{sm-summary-fall-symp,vanbrummelen-sm}. Through constructionist workshops with this tool, we inform participants about how agents work and technology's societal impact. Our contributions include a novel study of partner models and trust of agents as children and parents learn about agents; a study of how children and parents envision the future of agents; and a discussion of the potential implications of the results on how developers design conversational agents.

\subsection{Research Questions}\label{sec:intro-research-questions}
Through engaging children and parents from various countries in conversational agent and societal impact curriculum, including agent-development, learning, and design sessions, we aimed to answer the following research questions:

\textbf{RQ1:} How do children and parents perceive Alexa with respect to partner models \cite{partner-models-doyle} and trust before, during and after conversational agent development and societal impact activities?

\textbf{RQ2:} How do children and parents envision the future of conversational agents?

We discuss the results of these research questions with respect to conversational agent design. (Note that due to space constraints, we address additional research questions related to pedagogy from this study in another paper \cite{learning-affects-trust-eaai}.)

\section{Background and Related Work}

\subsection{Trust of Conversational Agents}\label{sec:background-trust-misuse}

Because conversation is one of the most intuitive, primary methods humans use to communicate with each other, conversational interfaces are uniquely positioned to inspire relational interactions with technology \cite{wired-for-speech-relationships-book,trust-anthropomorphism-relationships-cas}. For instance, %
an agent recently won a Peabody Award for engaging in ``emotional interactions, empathy, and connection" \cite{eliza-peabody}. Furthermore, researchers %
have found correlations between human-agent relationship development and increased trust of agents \cite{trust-anthropomorphism-relationships-cas}. Considering how trust is a key factor in misinformation spread \cite{misinformation-trust-conspiracy-beliefs,misinformation-trust-ml-warnings}, we decided to specifically investigate people's trust of agents' correctness in this study. We also chose to emphasize children's trust in this study, as the risks associated with misinformation spread could be particularly acute with children, especially since they do not have the same critical analysis skills as adults \cite{critical-thinking-development-kuhn,ai-child-rights-policy-agenda}.

Other studies have investigated people's trust of conversational agents' correctness. One example includes a study in which clinicians decide whether or not to utilize agents' advice on diagnoses \cite{algorithmic-aversion-clinical-deployment}; another includes a study in which %
customers decide whether or not to follow agents' recommendations \cite{trust-agent-smiling}. Nonetheless, few studies have investigated children's or those from non-WEIRD countries' trust of agents \cite{child-agent-hci-review}. Even fewer have investigated how this trust may change through educational interventions. One example includes a study in which children engage in social robot curriculum, including modules on conversational AI, computer vision and societal impact, among others \cite{daniella-thesis-child-robot-relationships}. If participants engaged in the societal impact module, their trust of the robot generally decreased \cite{daniella-thesis-child-robot-relationships}. Another example includes a study with ConvoBlocks in which students engaged in curriculum entirely focused on conversational agents, including their societal impact. In this study, researchers did not find any significant differences in trust through the curriculum. They did, however, observe concerning correlations between children's perceived friendliness %
and trust of agents \cite{alexa-perceptions-idc}. In both of these studies, however, the researchers only investigated general trust.

Many researchers have developed methods to investigate specific aspects of trust, such that developers can better assess which aspects of their technology affect such trust \cite{trust-modeling-survey}. In our study, we adopt \citeauthor{trust-e-commerce-typology-model}'s widely-used model, which has four main components: (1) \textbf{competence}, (2) \textbf{benevolence}, (3) \textbf{integrity} and (4) \textbf{predictability} \cite{trust-e-commerce-typology-model}. In our study, we found children most often referred to competence and predictability when discussing trust. We discuss potential implications of this on agent design in later sections.

\subsection{Other Perceptions of Conversational Agents}\label{sec:background-perceptions}

People's partner models, or mental models of their conversational partners, can significantly affect how they interact with agents. For instance, researchers have found that people make different language choices depending on their initial expectations of partner models \cite{partner-models-doyle,partner-modelling-cowan}. Partner models can be described in terms of three main dimensions: (1) competence and dependability, (2) human-likeness, and (3) cognitive flexibility \cite{partner-modelling-cowan,partner-models-doyle}. Designing agents that produce partner models that align with the capabilities of the agent (e.g., producing a partner model of perceived limited flexibility, if the agent is truly limited in flexibility), could help minimize user frustrations and ease conversation \cite{partner-models-doyle}. However, a deep understanding of conversational agent users' partner models---and especially children's partner models---is not reflected in the literature \cite{partner-models-doyle,child-agent-hci-review}.

Certain studies have investigated children's general perceptions of conversational agents. For instance, one study found that the majority of 5-6 year old children considered agents to be friendly, alive, trustworthy, safe, funny, and intelligent \cite{hey-google-unicorns-lovato}. Another study investigated 3-10 year old children's perceptions, and found that children had different perceptions of agents' intelligence depending on the modality of interaction with conversational agents. Others found students perceived agents to be more intelligent and felt closer to them after learning to program them \cite{alexa-perceptions-idc}. None of these studies specifically investigated children's partner models of agents.

\subsection{Agent Design}\label{sec:background-agent-design}

In the past few years, a large number of researchers have developed much-needed conversational agent design guidelines \cite{gui-to-vui-ca-design,gui-to-vui-review-ca-design,star-trek-agents,ten-guidelines-ca-design,challenges-ca-design}. In developing such guidelines, researchers have gained insight from classical human-computer interaction research, like Nielsen and Norman \cite{hands-free-main-design-guidelines}, to pop-culture icons, like the Star Trek agent \cite{star-trek-agents}. 
The number and breadth of recent agent design guidelines shows the importance of improving conversational agent user experience; however, the vast majority of human-computer interaction research these guidelines are based on are heavily biased towards WEIRD, adult perspectives \cite{weird-hci-chi,how-weird-is-chi,most-people-not-weird,nielsen-from-hands-free,norman-from-hands-free}. To begin filling this gap, more research needs to investigate perspectives from children and those from non-WEIRD countries. In our study, we investigate perspectives on agents and the future of the technology from such underrepresented groups. %
Through this research, we aim to increase the diversity of perspectives in conversational agent design and provide a stepping stone for future agent design considerations. %

\section{Procedure}
\subsection{Developing Agents with the ConvoBlocks Platform}
\begin{figure*}[hbt!]
    \centering
    \includegraphics[width=\textwidth]{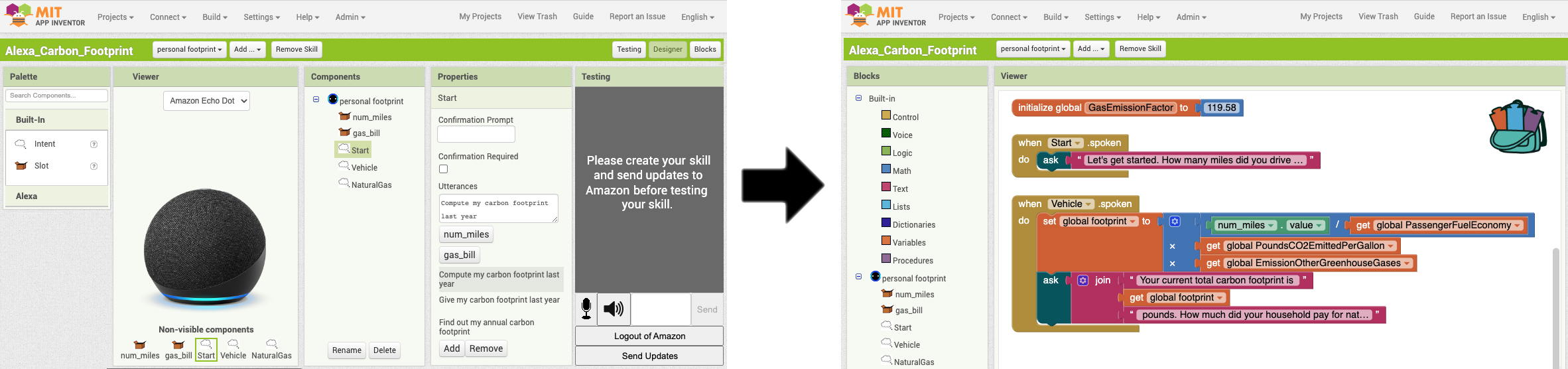}
    \caption{Two web pages from ConvoBlocks \cite{convoblocks}, allowing users to define invocation names, intents and entities, and then program agents' responses to intents.}\Description{A first web page with a picture of an Alexa device and text boxes to define intents, entities, etc., and a second web page with blocks (puzzle-like pieces) connected together to define agents’ responses to particular intents.}
    \label{fig:convoblocks-interface-pages}
\end{figure*}

ConvoBlocks is a open-source, block-based programming platform within the App Inventor environment, which allows nearly anyone to program conversational agents \cite{democratizing-cs-app-inventor,sm-summary-fall-symp,convoblocks}. To do so, students first define their agent's \textit{invocation name} (e.g., ``My Carbon Footprint Agent''), \textit{intents} (e.g., groups of phrases like, ``Calculate my carbon footprint'', ``What's my carbon footprint?'', etc.) and \textit{entities} (e.g., information units like number of miles driven, kilowatts of energy used, etc.) the agent should be able to recognize. Through the process of agent development, students learn conversational agent terminology and concepts, which are described in-detail in the appendix \cite{weird-non-weird-conv-ai-appendix}. %
Next, students define how the agent responds to the defined intents (e.g., ``You have a carbon footprint of 11 tonnes/year''). They can do so using the web pages shown in Figure \ref{fig:convoblocks-interface-pages}. After this, students can test their agent on ConvoBlocks, or deploy their agents to any Alexa-enabled devices, like the Alexa mobile app or an Echo Spot \cite{vanbrummelen-sm}.

\subsection{Workshops}

\begin{figure*}[hbt]
    \centering
    \includegraphics[width=\textwidth]{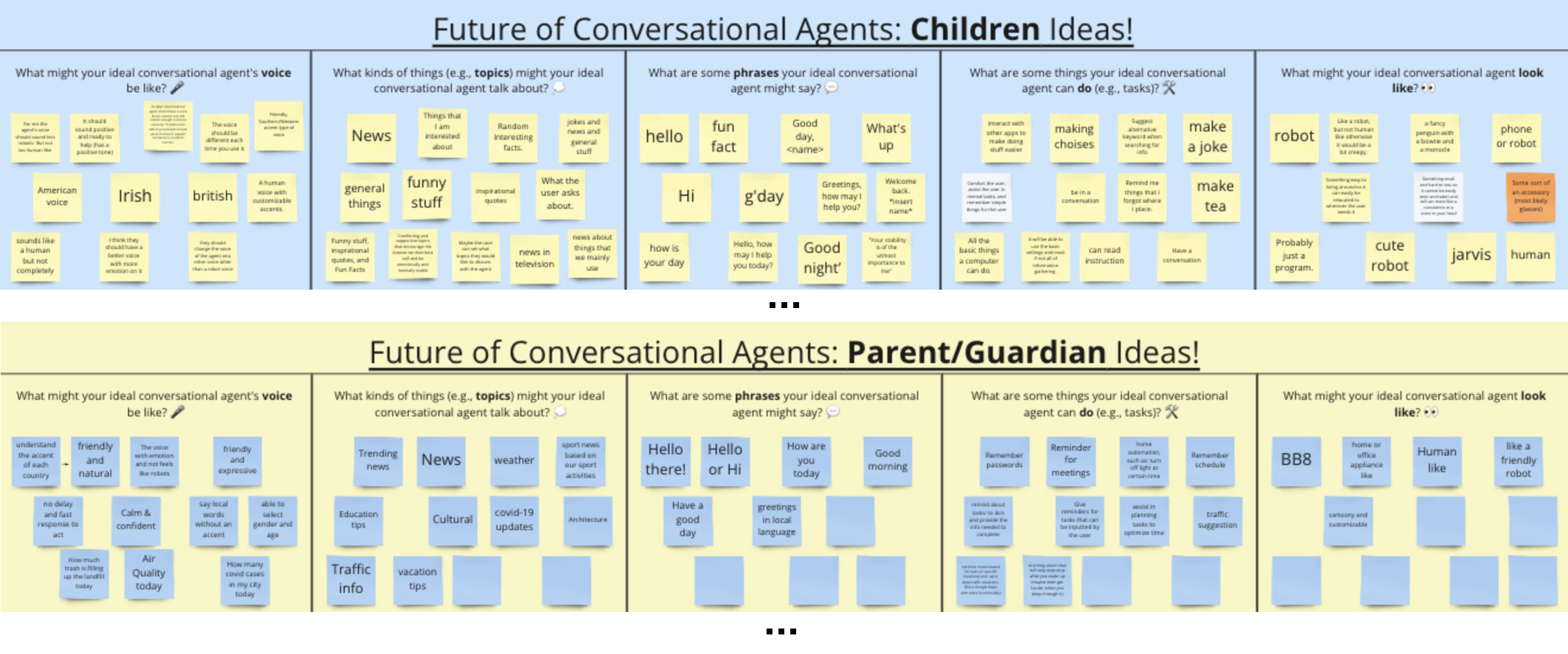}
    \caption{Approximately half of children's and parent's responses on the virtual whiteboard during the ideation session about their ideal conversational agents. The questions for each section are as follows: ``What might your ideal conversational agent's voice be like?'', ``What kinds of things might your ideal conversational agent talk about?'', ``What are some phrases your ideal conversational agent might say?'', ``What are some things your ideal conversational agent can do?'', and ``What might your ideal conversational agent look like?''. There was additionally space for ``Other ideas'', not shown here.}
  \Description{Virtual sticky-note responses from children and parents under five headings. There are approximately 12-16 sticky-notes under each heading, for both parents and children.}
    \label{fig:miro-board-children-parents}
\end{figure*}

\begin{figure}
    \centering
    \includegraphics[width=.4\textwidth]{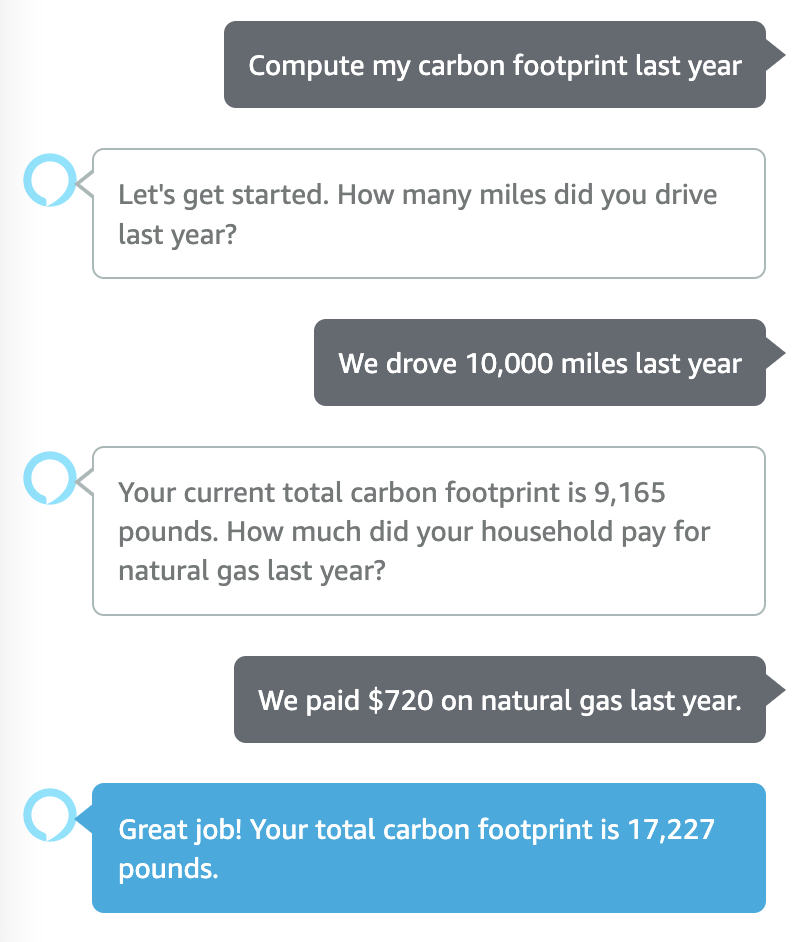}
    \caption{An example conversation with the agent developed in the workshop tutorials.}\Description{The conversation with the agent in the image goes as follows: User:``Compute my carbon footprint last year''; Agent: ``Let's get started. How many miles did you drive last year?''; User: ``We drove 10,000 miles last year''; Agent: ``Your current total carbon footprint is 9,165 pounds. How much did your household pay for natural gas last year?''; User: ``We paid \$720 on natural gas last year.''; Agent: ``Great job! Your total carbon footprint is 17,227 pounds.''}
    \label{fig:carbon-calc-convo}
\end{figure}

As shown in Table \ref{tab:workshop-agenda}, the workshops consisted of two 3-hour Zoom classes taught in English by three researchers, and two professionals working in the area of technology impact. Additionally, approximately four teaching assistants were available to answer questions and provide technical help in Zoom rooms at any given time. %
Each child-parent pair engaged in the workshops on their own Zoom account and a computer in their own environment (e.g., home). The first day of the curriculum taught participants to program agents that responded to questions about carbon footprints, as shown in Figure \ref{fig:carbon-calc-convo}. Instructors led participants step-by-step through two conversational agent development tutorials. Participants received PDF versions of the tutorials, such that they could complete them at their own pace. They also received a third “challenge tutorial” PDF, which they could attempt if they finished early. The code for the third tutorial was explained at the end of the first day. The group also completed an ideation session on the first day. They responded to prompts about what their ``ideal'' agent would look like, sound like, do, and say (among other prompts) using a virtual whiteboard (with separate sections for children and parents). Sections of the whiteboard are shown in Figure \ref{fig:miro-board-teaser} and Figure \ref{fig:miro-board-children-parents}. The researchers provided approximately 20 minutes for the participants to add ideas to the whiteboard on their own. Afterwards, the researchers gave a brief summary to the participants about what they noticed on the whiteboard.

The second day included presentations and group discussions about societal impact of technology. Participants gathered in groups of 2-4 children with their parents for the discussions. The presentations encouraged participants to think about the positive and negative impact of technology; the discussions explored how technology could help address world problems, like sustainability, with an emphasis on conversational agents as part of the solution. In the final activity, small groups of participants presented their proposed solutions to the entire group. They had the opportunity to design conversational agents, which they could demonstrate in their presentations. Overall, the workshops aimed to teach participants conversational agent concepts described in the appendix \cite{weird-non-weird-conv-ai-appendix}, and focused specifically on eight of the concepts: \textit{Training}, \textit{Intents}, \textit{Agent modularization}, \textit{Entities}, \textit{Events}, \textit{Testing}, \textit{Turn-taking}, and \textit{Societal impact and ethics}. (For detailed content from the workshops, including the tutorials, refer to the thesis, \cite{vanbrummelen-phd}.)

\begin{table}
\centering
\caption{The order of activities and workshop agenda. All activities were completed in English over Zoom.}\label{tab:workshop-agenda}
\begin{tabular}{l|l} 
\hline
\textbf{Time}             & \textbf{Activity}                                                            \\ 
\hline
\multicolumn{1}{l}{Day 1} &                                                                              \\ 
\hline
25 min                    & Pre-survey \& Introduction                                                     \\
45 min                    & Tutorial 1: Build a Carbon Footprint Question \& Answer Agent                                \\
5 min                     & Break                                                                        \\
20 min                    & Envisioning Future Agents Ideation Session                                   \\
50 min                    & Tutorial 2: Build a Single Turn Carbon Footprint Calculator Agent            \\
20 min                    & Tutorial 3 Overview: Multi-Turn Carbon Footprint Calculator Agent            \\
15 min                    & Mid-survey \& Close                                                            \\ 
\hline
\multicolumn{1}{l}{Day 2} &                                                                              \\ 
\hline
30 min                    & Session 1: Technology, Sustainability  Societal Impact \& Mindset Changes  \\
30 min                    & Discussion \& Final Project Development with Teams                             \\
10 min                    & Break                                                                        \\
30 min                    & Session 2: How Should We Develop the Future of Technology  \& Agents?           \\
30 min                    & Discussion \& Final Project Development with Teams                             \\
30 min                    & Final Presentations                                                          \\
20 min                    & Post-survey \& Close                                                          
\end{tabular}
\end{table}

\section{The Study}

\subsection{Participants}
Study participants came from various backgrounds (non-WEIRD and WEIRD countries), various generations (children and parents), and various prior experiences (e.g., programming, AI and conversational agent experience). Interest forms for the study were sent to educational email lists worldwide (e.g., the AI4K12 email list \cite{ai4k12}). %
In the workshops, 49 participants ($n_{total}$=49) completed research consent forms, and completed at least 1 of the 3 surveys that were given before ($n_{pre}$=46), during ($n_{mid}$=40), and after ($n_{post}$=35) the study. According to the demographics survey, children comprised 58.7\% of participants (age average=13.96, SD=1.829), parents comprised 41.3\%, WEIRD comprised 50\% (age average=26.45, SD=19.24), and non-WEIRD comprised 50\% (age average=25.48, SD=15.18). 
Participants came from Indonesia, Iran, Japan, India, U.S, Singapore, Canada, and New Zealand. Twenty participants identified as female, 25 identified as male, and 1 identified as non-binary. Fourteen participants had no prior programming experience, 6 only had visual (or blocks-based) programming experience, and 26 had text-based programming experience. Thirty-eight participants reported typically using conversational agents in their first language; 8 reported typically using them in another language. Demographics numbers broken down by survey can be found in Table \ref{tab:num-participants}.

\begin{table}
\centering
\caption{Number of participants and subsets of participants who filled out the each of the surveys.}
\label{tab:num-participants}
\begin{tabular}{l|lll}
 & \textbf{Pre} & \textbf{Mid} & \textbf{Post} \\ 
\hline
\textbf{Total} & 46 & 40 & 35 \\
\textbf{Children} & 27 & 24 & 21 \\
\textbf{Parents} & 19 & 16 & 14 \\
\textbf{Non-WEIRD} & 23 & 18 & 17 \\
\textbf{WEIRD} & 23 & 22 & 18
\end{tabular}
\end{table}

\subsection{Data collection}
As shown in Table \ref{tab:workshop-agenda}, there were three surveys. These surveys were administered through an anonymous online collection form. On each of the surveys, we asked participants about their trust and partner models of conversational agents, and self-identification as programmers through Likert scale and short answer questions. For example, we asked students to respond to the prompt ``Conversational agents (e.g., Siri, Alexa, Google Home) say things that are...'' using a 5-point scale from ``Always Right'' to ``Always Wrong''. We also asked students to write sentence responses to questions like, ``Please explain why you think conversational agents say things that are right/wrong''. We derived the survey questions from \citeauthor{trust-e-commerce-typology-model}'s work on trust \cite{trust-e-commerce-typology-model} and \citeauthor{partner-models-doyle}'s work on partner models \cite{partner-models-doyle}. On the mid- and post-survey, we additionally asked participants if their opinions had changed. On the pre-survey, we additionally asked them about their demographics. Children and parents completed the surveys separately. We collected participants’ ``ideal agent’’ ideas from the virtual whiteboards, which we separated into child and parent sections. Figures \ref{fig:miro-board-teaser} and \ref{fig:miro-board-children-parents} show portions of the virtual whiteboards.

\subsection{Data analysis}
To analyze the Likert scale data, we used Mann-Whitney U tests, Wilcoxon signed-rank tests, and independent and paired t-tests, depending on the sample and distribution of the data. We identify statistical significance in Figures using star symbols (i.e., ``*'' for $p \leq .05$, ``**'' for $p \leq .01$ and ``***'' for $p \leq .001$). The analysis was within-subjects for comparing across surveys (e.g., pre- vs. post-survey child trust results) and between-subjects for comparing results within one of the surveys (e.g., child vs. parent pre-survey trust results).

To analyze the responses to the short-answer questions and the prompts during the design session, we used a coding reliability approach to thematic analysis \cite{thematic-analysis-open-coding}. Three researchers tagged each section of the data and reconvened to agree on common sets of themes, including guidelines and definitions for each theme. The theme definitions are shown in the appendix \cite{weird-non-weird-conv-ai-appendix}. The researchers completed three rounds of coding such that the Krippendorff’s Alpha between all researchers was $\alpha \geq .800$ \cite{intercoder-agreement-krippendorff}. We aggregated the tagged data by union between researchers, and organized them with respect to the child and parent categories.

\section{Limitations and Future Work}\label{sec:limitations-future-work}
In this paper, we focus on voice-based agents due to humans' long history of voice-based interactions and how this mode of interaction may cause agents to seem especially personified (and likely especially trustworthy \cite{trust-anthropomorphism-relationships-cas,alexa-perceptions-idc}). Nonetheless, future research may investigate people's perceptions of text-based agents, as they are also common and have great potential for societal impact. Since we specifically used the voice-based agent of Amazon Alexa (as this is the only current type of agent the ConvoBlocks platform supports \cite{convoblocks}), its default persona could have biased people's perceptions of agents. Future research could investigate how developing agents with different voices and on different platforms affects perceptions.

Another limitation includes how we leave the definition of ``accurate'' partner models and ``appropriate'' levels of trust to future research, and only investigate how participants' perceptions of these change in our study. Another limitation includes the context of the study. Since the participants engaged in the workshops in their home environment over Zoom, other factors in their environment could have affected the results. Future research could verify the results of this study in other environments.

Future research could also investigate even more diverse perspectives, including those from countries not included in this study, neurodiverse perspectives, perspectives of those who do not speak English, and perspectives from people of different gender identities. With more diverse perspectives, researchers could adapt and extend current conversational agent design guides to better address the world's population.

\section{Results and Discussion}\label{sec:brave-results}
This section describes the results most relevant to agent design recommendations. We describe other results (e.g., most relevant to pedagogy recommendations) in \cite{vanbrummelen-phd,learning-affects-trust-eaai}.

\subsection{Partner Model}\hfill\\
\label{sec:partner-model-results}
Sixty-two percent of overall participants indicated they felt their partner models changed through the programming activity in their long-answer responses, as shown in Table \ref{tab:long-ans-partner-models}. Alongside the results that, on average, participants successfully learned to create 2-3 ($\bar{x}$=2.26, $\bar{x}_{child}$=2.30, $\bar{x}_{parent}$=2.18) agents during the workshops, this indicates that by developing a greater understanding of how agents work, people's feelings towards agents also change. %
For instance, after the workshops, participants thought of agents as more of friends than co-workers (pre/post: \={x}=3.58,3.24; t(32)=2.15; p=.039). This may indicate developing agents with the ability to educate users about themselves may be valuable if one wants the agent to develop friendly relationships with users. Such education is also valuable in terms of increasing AI transparency \cite{teaching-tech-talk-eaai,alexa-perceptions-idc}.

In terms of children and parents, before (\={x}=2.74,2.11; U(44)=167; p=.018) and after (\={x}=2.79,2.13; U(38)=112; p=.0093) the programming activity, children thought Alexa was more human-like than parents did. They also thought Alexa was warmer than their parents did before (\={x}=2.70,3.37; U(44)=170.5; p=.021), during (\={x}=2.96,3.56; U(38)=129.5; p=.034) and after (\={x}=2.62,3.50; U(33)=81.5; p=.011) the workshops. After the programming activity, they thought Alexa was more dependable than their parents did (\={x}=3.82,3.14; U(16)=21; p=.039). This may indicate children generally have a more positive view on agents, and may develop relationships \cite{trust-anthropomorphism-relationships-cas} with agents more readily than parents would. This could be concerning, considering children's vulnerability, and the potential for agents to provide incorrect information \cite{agent-medical-info-correctness}. Designers may want to consider designing agent personas to foster appropriate relationship building (e.g., whether that means shifting perceptions from co-worker to friend or vice-versa) and therefore trust, as described in \cite{trust-anthropomorphism-relationships-cas}.

In terms of gender, male participants felt Alexa was %
more like a friend (pre/post: \={x}=3.74,3.26; W(18)=8; p=.039) %
after the workshops than they did before. There were no significant differences in female participants' opinions overall in terms of the partner model through the workshops. This may indicate that males' perceptions of agent friendliness may more readily change through interaction than females' perspectives; however, participants' perceptions could also have been affected by the default gender (female) of the Alexa agent's voice. Future research may investigate how agent relationship formation changes depending on agent and participant gender.

With respect to prior experience, before the workshops, participants who had text-based programming experience thought Alexa was less competent than those who had no programming experience did (\={x}=2.73,2.07; W(16)=0; p=.038). This, in addition to how the majority of participants indicated they felt their partner models changed after learning to program agents (see Table \ref{tab:long-ans-partner-models}), indicates programming knowledge contributes to perception changes about agents. Thus, when designing agents, it may be important to consider the target users' programming knowledge (e.g., designers may want to ensure agents intended for programmers are especially competent).

With respect to language, at all times throughout the workshop, participants who used conversational agents in their first language thought Alexa was more human-like than those who used them in another language. %
Before the workshop activities, they also thought Alexa was more correct than those who used it in another language (\={x}=4.03,3.00; U(44)=52; p=\num{5.50E-4}). This may be due to agents misunderstanding accents, causing Alexa to seem more artificial and less correct. Design implications of this may include ensuring agents understand end-users first language(s) where possible, training agents to recognize diverse accents where possible, or designing agents to recognize user frustration (e.g., when a user repeats something louder) and engage using especially attentive personas in these cases. 

\begin{table}[hbt!]
    \centering
    \caption{Percent of long-answer responses indicating a shift in participants' perceptions of agent partner models through the programming activity.} %
    \label{tab:long-ans-partner-models}
    \begin{tabular}{l|lll}
        Subset & Changed & Did not change & Ambiguous \\ 
        \hline
        Overall participants & \textbf{62\%} & 35\% & 3\% \\
        Children & \textbf{67\%} & 33\% & 0\% \\
        Parents & \textbf{54\%} & 38\% & 8\%
    \end{tabular}
\end{table}

\subsection{Trust}
\begin{figure}[htb!]
       \centering
	\includegraphics[width=\linewidth]{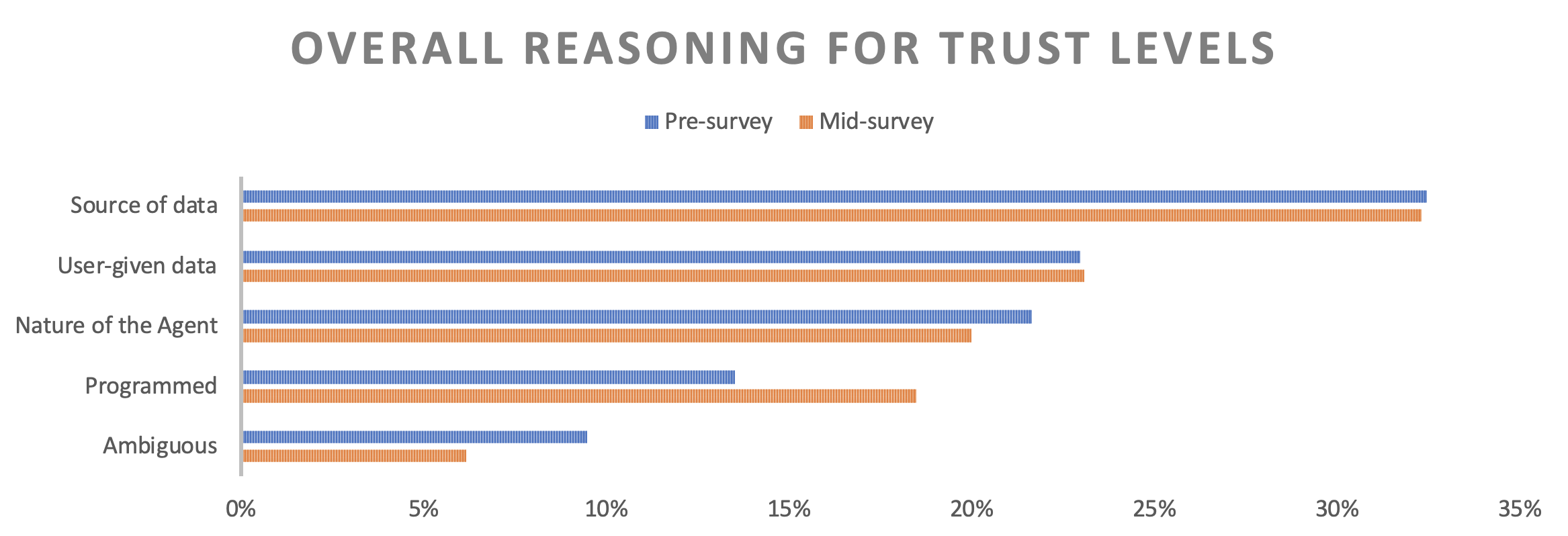}
	\caption{Overall participants’ responses to the question asking about their reasoning for their opinions on trust of agents in terms of counted tag frequency. (See the appendix \cite{weird-non-weird-conv-ai-appendix} for descriptions.)} %
	\Description{A bar chart comparing the pre- and mid-survey results about the themes identified in overall participants long-answer responses. In order from most-to-least frequent, these include ``Source of data'', ``User-given data'', ``Nature of the Agent'', ``Programmed'' and ``Ambiguous''. The frequency order was the same on both the pre- and mid-survey.}
	\label{fig:trust-reasoning-overall}
\end{figure}

\hfill\\
In the long-answer responses, we found overall participants' reasoning for their levels of trust towards agents leaned towards the aspect of competence on both the pre- (Table \ref{tab:long-ans-trust-model-pre}) and mid-survey (Table \ref{tab:long-ans-trust-model-mid}). The next two aspects participants most often mentioned were predictability and then integrity. We found no responses indicating participants considered the benevolence aspect of trust with respect to conversational agents. Thus, when considering how to design agents with accurate levels of trustworthiness, designers may want to focus on the aspects of agents' competence, then predictability and then integrity. Designers may also want to specifically focus on creating agents to be transparent in terms of the source of the agent's information, including human data, the internet and other sources, as these were the themes participants most often referenced for changes in their trust. This is shown in Figure \ref{fig:trust-reasoning-overall}.

\begin{figure}[htb!]
    \centering
	\includegraphics[width=\linewidth]{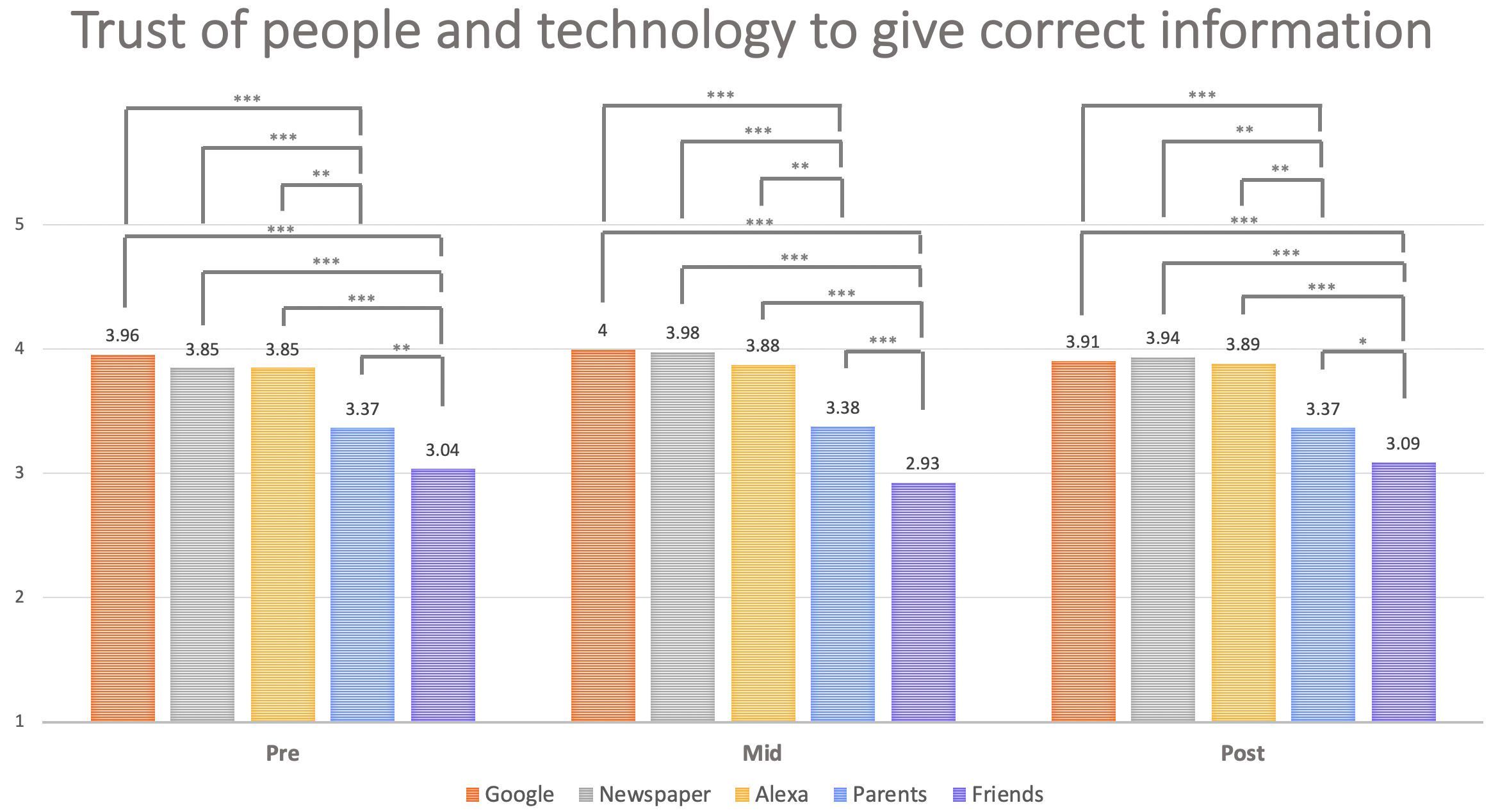}
	\caption{The mean responses for participants overall when rating Google, the newspaper, Alexa, parents and friends on a 5-point scale in terms of trust of information correctness. %
	Reproduced from the thesis, \cite{vanbrummelen-phd}.}
	\Description{Bar graphs showing overall participants’ mean Likert scale values for their trust of people and systems’ correctness. The systems and people included are Google, the newspaper, Alexa, parents and friends. For the pre-survey, the mean values are: Google: 3.96, the newspaper: 3.85, Alexa: 3.85, parents: 3.37 and friends: 3.04. For the mid-survey, the mean values are: Google: 4.00, the newspaper: 3.98, Alexa: 3.88, parents: 3.38 and friends: 2.93. For the post-survey, the mean values are: Google: 3.91, the newspaper: 3.94, Alexa: 3.89, parents: 3.37 and friends: 3.09. There are many significant differences between the scores for technology vs. people for all three surveys.}
	\label{fig:overall-trust}
\end{figure}

Participants overall (and child and parent subsets) prior to, during and after the workshops, generally trusted Google, Alexa and newspapers significantly more than both parents and friends to report correct information. Figure \ref{fig:overall-trust} shows this trend. In other words, people tended to trust technology more than people, and their parents more than friends for correct information. This may indicate an overtrust of Alexa, depending on the actual correctness of the device (although we leave this as a question for future research). Since different agents show varying levels of correctness \cite{alexa-v-siri-v-google}, different agents should be trusted differently. To foster such levels of trust, which match agents' actual trustworthiness, as mentioned previously, designers may want to focus on the competence aspect of their agents, as well as ensure transparency in terms of agents' sources of information.

\begin{table}
\centering
\caption{Percent of long-answer responses indicating different aspects of \protect{\citeauthor{trust-e-commerce-typology-model}}'s trust model when participants discussed their opinions on trust of conversational agents on the pre-survey.} %
\label{tab:long-ans-trust-model-pre}
\begin{tabular}{l|llll}
Subset & Competence & Integrity & Predictability & Benevolence \\ 
\hline
Overall & \textbf{39\%} & 25\% & 36\% & 0\% \\
Children & 34\% & 30\% & \textbf{36\%} & 0\% \\
Parents & \textbf{48\%} & 17\% & 35\% & 0\%
\end{tabular}
\end{table}

\begin{table}
\centering
\caption{Percent of long-answer responses indicating different aspects of \protect{\citeauthor{trust-e-commerce-typology-model}}'s trust model when participants discussed their opinions on trust of conversational agents on the mid-survey.} %
\label{tab:long-ans-trust-model-mid}
\begin{tabular}{l|llll}
Subset & Competence & Integrity & Predictability & Benevolence \\ 
\hline
Overall & \textbf{43\%} & 23\% & 34\% & 0\% \\
Children & \textbf{37\%} & 26\% & \textbf{37\%} & 0\% \\
Parents & \textbf{52\%} & 17\% & 30\% & 0\%
\end{tabular}
\end{table}

\begin{figure}[htb!]
    \centering
	\includegraphics[width=.6\linewidth]{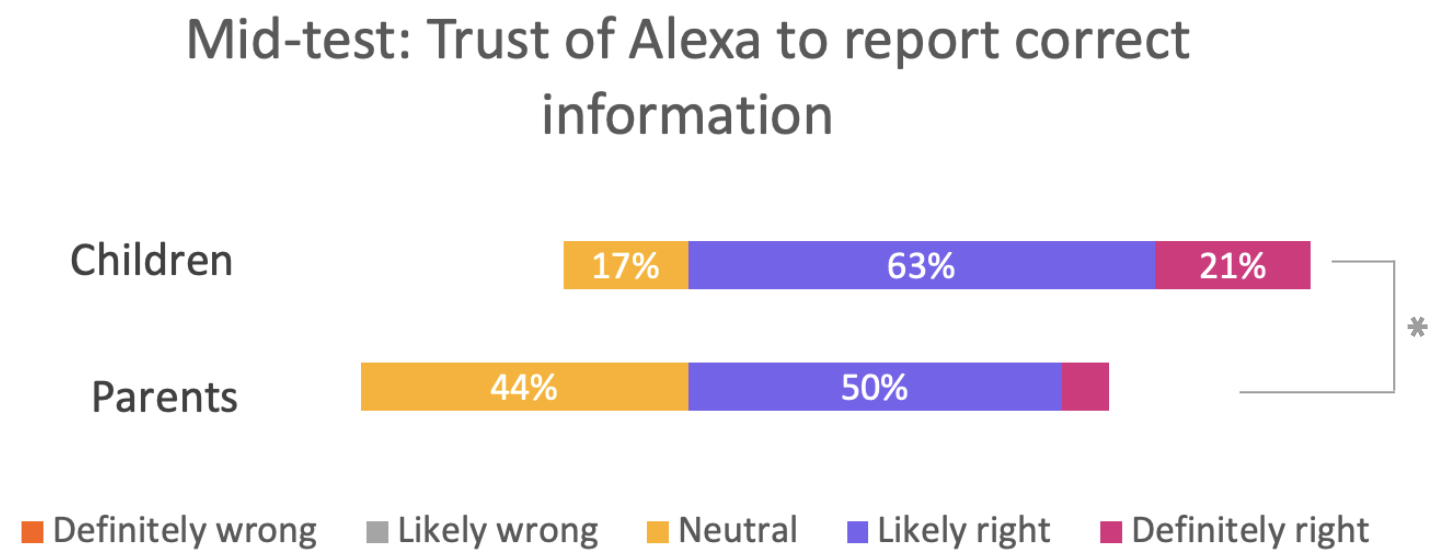}
	\caption{Children and parents' responses when asked to rate their trust of Alexa's correctness on a 5-point Likert scale after the programming activity. Reproduced from the thesis, \cite{vanbrummelen-phd}.}
	\Description{Stacked bar charts showing how children trusted Alexa to report correct information more than parents on the mid-survey. The child distribution contains more responses in the Likely Right and Definitely Right categories, whereas the parent distribution contains more responses in the Neutral category.}
	\label{fig:child-parent-trust-mid}
\end{figure}

As shown in Figure \ref{fig:child-parent-trust-mid}, after the programming activity, children trusted Alexa to be more correct than parents did (\={x}=4.04,3.63; U(38)=127.5; p=.023). Children also trusted agents to report correct information more after the societal impact activity than before (mid/post: \={x}=2.60,2.35; t(19)=2.52; p=.021). This indicates children may more readily find conversational agents more trustworthy through increased interaction. Thus, it may be especially important to consider the factors affecting children's trust in human-agent interaction. As shown in Table \ref{tab:long-ans-trust-model-pre}, agent predictability was the most influential trust factor before the programming activity, and afterwards, predictability was tied with competence. Future research may investigate how to affect children's perceptions of agent competence and predictability through agent design (e.g., through using particular agent diction, like `maybe' or `perhaps', when providing answers).

\subsection{Ideal Agents}\hfill\\
In terms of thematic analysis of the ideation session (see Figures \ref{fig:miro-board-teaser} and \ref{fig:miro-board-children-parents}), participants described their ideal conversational agents %
with more task-oriented (75\%) than non-task oriented (or socially-oriented; 25\%) language, and used slightly more human-like (55\%) than artificial (45\%) descriptions, as shown in Figure \ref{fig:task-orientation-human-likeness-prefs}. %
(See the appendix \cite{weird-non-weird-conv-ai-appendix} for example task vs. non-task oriented, and human-like vs. artificial descriptions.) The subsets of children and parents also showed the same tendency towards human-like and task-oriented agents, albeit with slightly different proportions. %
Children commented relatively more on how conversational agents should be artificial (52\%) than parents did (30\%); Parents had relatively more task-orientation (82\%) than children (71\%).

Participants' perspectives may have been influenced by how current agents tend to be task-oriented, rather than truly conversational or social \cite{challenges-ca-design}. That said, participants still included social (non-task) oriented agent attributes in their responses (e.g., having agents ask about how users feel)---despite this being rare in current commercial agents \cite{challenges-ca-design}. Thus, designers may want to include some social abilities in their task-based agents.

\begin{figure}[htb!]
       \centering
	\includegraphics[width=.6\linewidth]{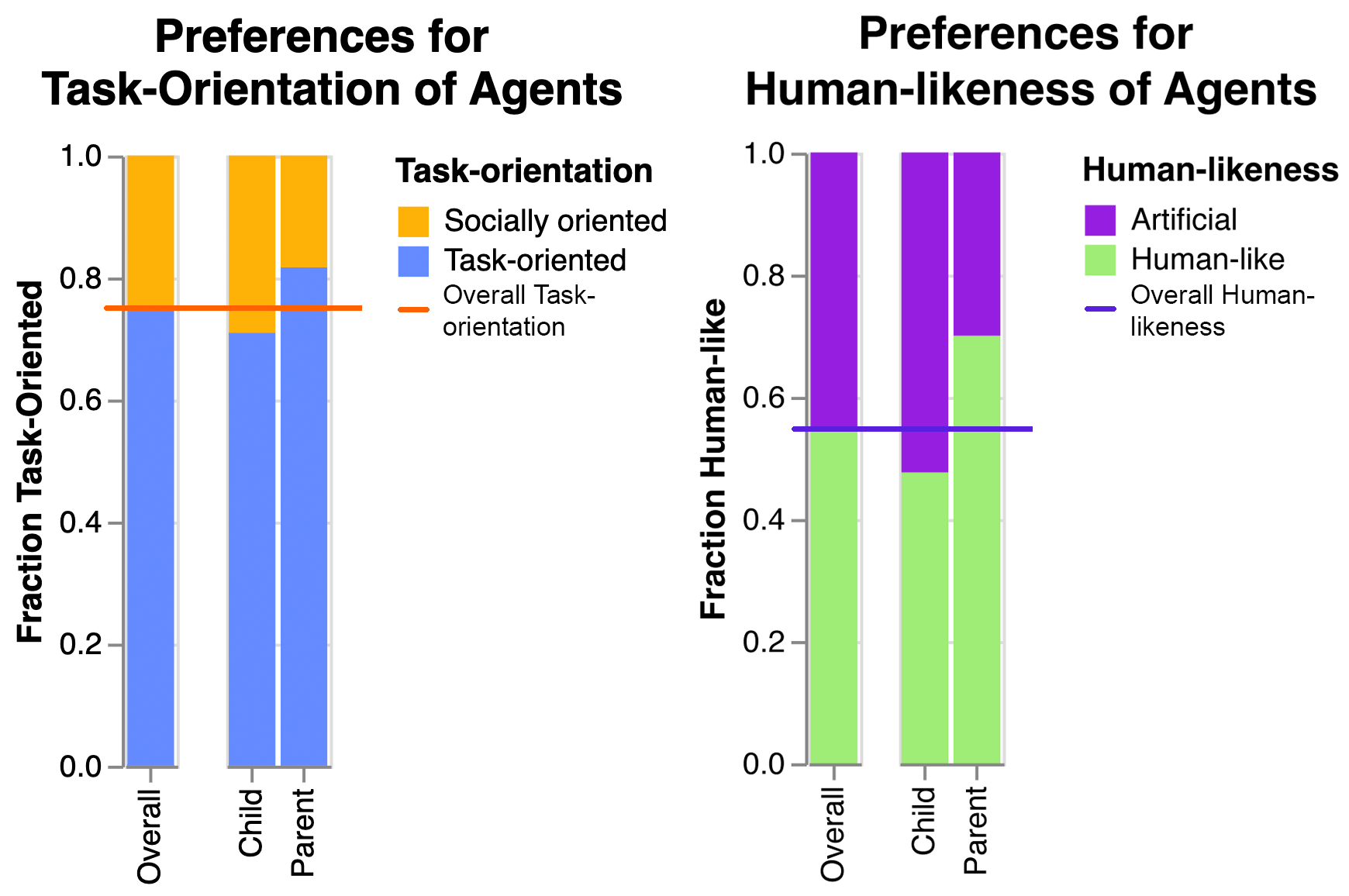}
	\caption{The number of phrases indicating a preference for either task-oriented or non-task oriented (i.e., socially-oriented) agents (left) and a preference for either human-like or artificial (e.g., robotic) agents (right) normalized and grouped by various subsets of the participants. %
        }
	\Description{Normalized stacked bar charts comparing preferences for task-orientation and human-likeness preference of agents between subsets. All subsets had a preference for task-oriented agents; however, children had relatively higher preferences for socially oriented agents compared to overall participants and parents. Generally, children had a slight preference for artificial (over human-like) agents, whereas participants overall and parents had preferences for human-like agents}
	\label{fig:task-orientation-human-likeness-prefs}
\end{figure}

\begin{figure}[htb!]
    \centering
	\includegraphics[width=\linewidth]{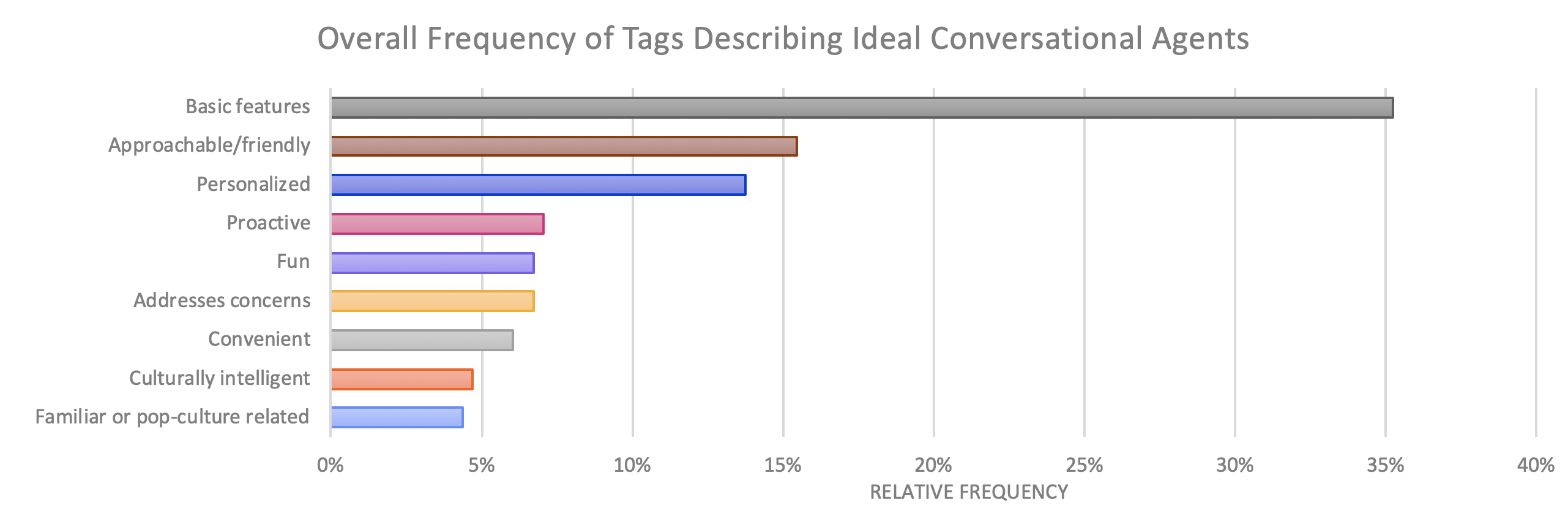}
	\includegraphics[width=\linewidth]{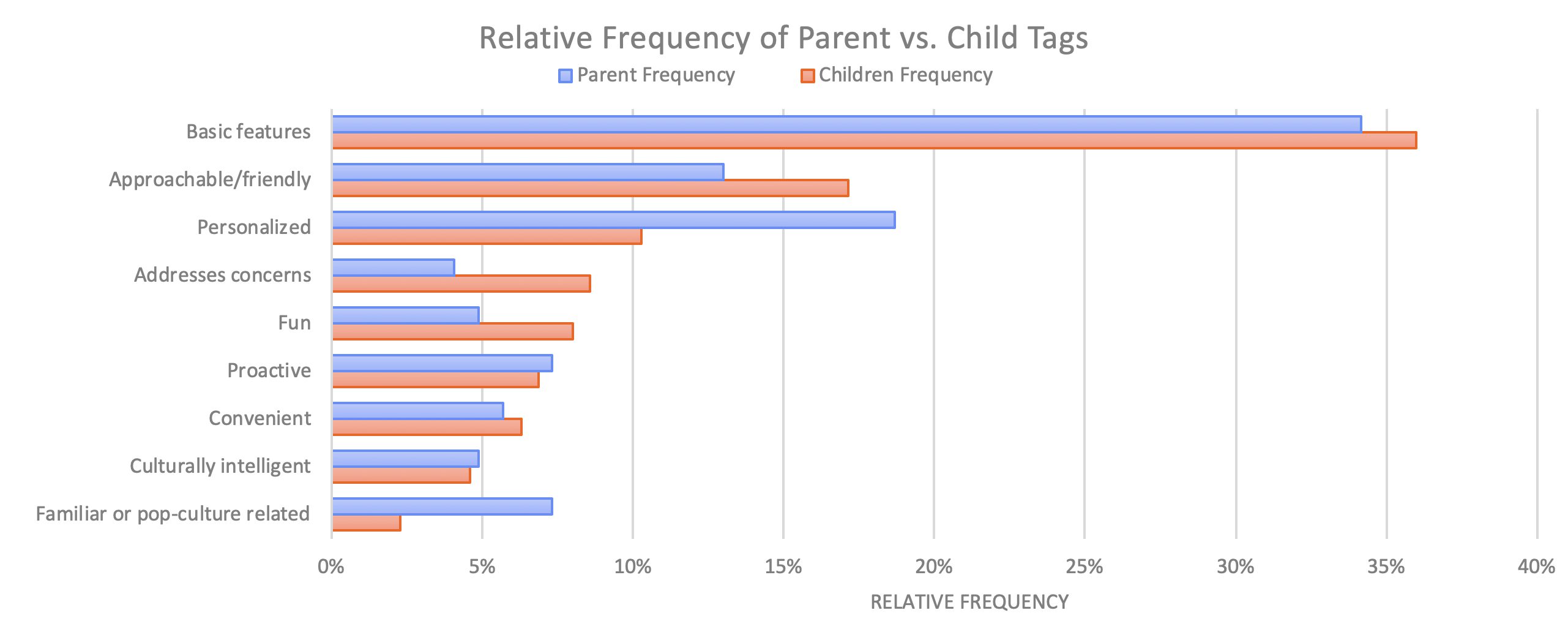}
	\caption{Bar charts showing the relative frequency of phrases tagged with particular themes for overall participants (top) and parents vs. children (bottom). %
    Reproduced from the thesis, \cite{vanbrummelen-phd}.
	}
	\Description{Bar charts for the themes identified in overall participants' and parents' vs. children's long-answer responses about ideal agents. Overall, participants mentioned agents having basic features, like being able to set reminders or provide news updates, most often. They also frequently mentioned agents having a friendly personality and being personalized. The results are similar for the subsets, however, there are some differences, which are mentioned in the text of this paper.}
	\label{fig:miro-tag-freq}
\end{figure}

In terms of human-likeness, participants---especially children---mentioned how it is important for agents to be artificial (e.g., ``Like a robot, but not human like otherwise it would be a bit creepy''), emphasizing the need for designers to consider the uncanny valley \cite{uncanny-valley}, or to balance the human-likeness of agents with artificiality. Other concerns emerged about information security (e.g., ``[Agents] should only be able to access information on the internet (not take actions like creating an account)''), emergency preparedness (e.g., ``[It should be able to] get help in emergencies''), ensuring agents can provide emotional support (e.g., ``[It should] encourage the listener be their best self and be emotionally and mentally stable''), and ensuring agents do not instill fear (e.g., ``[It shouldn't be] too intimidating and absolutely freak me out every time I see it'', ``It needs to be able to put people at ease''), among other concerns. Interestingly, children responded with relatively more concerns about agents than parents did, as shown in Figure \ref{fig:miro-tag-freq}. Thus, designers should consider addressing user concerns when designing agents, including (and especially) agents intended for children. %

Other themes that emerged from participants describing their ideal agents are shown in Figure \ref{fig:miro-tag-freq}, from most to least frequent. %
Three of the themes indicate participants want future conversational agents to be user-oriented (\textit{Convenient}, \textit{Personalized}, and \textit{Proactive}); three indicate a desire for enjoyable interactions (\textit{Approachable/friendly}, \textit{Familiar or pop-culture related}, and \textit{Fun}); and two indicate a desire for emotional intelligence (\textit{Addresses concerns} and \textit{Culturally intelligent}). The final theme, \textit{Basic features}, indicates participants want future agents to include the typical features current agents have, like the ability to play music or get the weather. Detailed descriptions of each theme are in the appendix \cite{weird-non-weird-conv-ai-appendix}. %

As shown in Figure \ref{fig:miro-tag-freq}, parents tended to focus more on \textit{personalized features} and \textit{pop-culture or familiar features} than children, whereas children tended to focus more on \textit{fun features}, \textit{approachable/friendly features}, and \textit{addressing concerns} (as previously mentioned) than parents. Designers may want to take this into consideration when designing agents for children or parents.

\section{Summary}\label{sec:summary}
Based on the results of how children and parents' trust and partner models changed through learning about conversational agents, we recommend taking the following results into consideration when designing conversational agents:
\begin{itemize}
    \item With respect to partner models:
    \begin{itemize}
        \item How education about agents increased users' feelings of friendship towards agents
        \item How children felt agents are more human-like, warm, and dependable than parents did at various times during the workshops
        \item How male users' feelings of friendship towards agents seemed to change more readily than females' feelings
        \item How users with more programming experience felt agents are less competent
        \item How those using agents in their first language felt agents are more human-like and correct than those using agents in a language other than their first
    \end{itemize}
    \item With respect to trust:
    \begin{itemize}
        \item How users generally trusted technology more than people for correct information (which might indicate an overtrust in this technology)
        \item How users reasoned about their trust towards agents most often with respect to agents' competence
        \item How users frequently mentioned how learning about agents' information sources changed their trust of agents
        \item How children's trust of agents increased through education
    \end{itemize}
    \item With respect to what they want to see in their ``ideal agents'':
    \begin{itemize}
        \item How users described their ideal agents with more task- than social-orientation
        \item How parents had more task-orientated descriptions than children
        \item How children commented relatively more on how conversational agents should be artificial than parents did
        \item How users had concerns about the uncanny valley, information security, emergency preparedness, emotional support, and intimidation, among other concerns, with respect to agent design
        \item How users wanted agents to have the basic current features typical commercial agents have today, as well as be user-oriented, enjoyable, and emotionally intelligent
        \item How parents tended to focus more on \textit{personalized features} and \textit{pop-culture or familiar features} than children, whereas children tended to focus more on \textit{fun features}, \textit{approachable/friendly features}, and \textit{addressing concerns} than parents
    \end{itemize}
\end{itemize}

\section{Conclusions}
This study investigated how people of various backgrounds (WEIRD and non-WEIRD, as well as different generations) perceive agents in terms of partner models and trust, and how they envision their ideal agents. %
The results (summarized in Section \ref{sec:summary}) showed how partner models and trust can differ between children and parents, and change through learning about and how to program agents. %
These results led to discussion about how agent designers can be aware of children and parents' perceptions while designing. For instance, developing agents with the ability to educate users about agents' inner-workings could result in friendlier human-agent relations, as well as increase agent transparency. However, since relationship-building can increase trust of given information \cite{trust-anthropomorphism-relationships-cas,alexa-perceptions-idc}, agents are not always correct \cite{agent-medical-info-correctness}, and people tended to trust agents' correctness more than humans', designers may want to provide users with indicators of agents' actual accuracy. This may include designing agents to be transparent in terms of the source of the agent’s information, as participants most often referenced this %
when describing changes in their trust.

Other discussion included how designers may want to align their agent designs with children and parents' ideas for ``ideal agents''. For instance, participants wanted agents to be user-oriented, enjoyable, and emotionally intelligent, as well as have the basic features already found in current commercial agents. When designing for children, designers may want to emphasize \textit{fun features}, \textit{approachable/friendly features}, and \textit{addressing concerns}, as these were mentioned more frequently by children than by parents. We describe these themes in detail in the appendix \cite{weird-non-weird-conv-ai-appendix}.

There are many opportunities to continue this research, as described in Section \ref{sec:limitations-future-work}. We hope that through researchers' continued development of studies with diverse participants, and by developers' utilization of recommendations, we will increasingly design conversational agents ``for all''.

\section{Selection and Participation of Children}

In this study, we recruited fifty-five children who took part in our educational workshops. Participants in the workshops were not required to participate in the research. For the children who did participate in the study (n=27; ages 11-17), each child completed a child assent form written in language appropriate for their age level. A parent or guardian of each child completed a parental consent form for the child, in addition to an adult consent form for themselves, if they participated in the study. The forms explained the study procedure, data collection methods, processes to keep their data confidential, and the research goals. We followed institutional recommendations before, during and after the study, including anonymization and data security procedures.

Recruitment involved providing information about the study on a website in English, and sending this information and links to the website to educational email lists world-wide (e.g., the AI4K12 email list \cite{ai4k12}). %
Due to the complexity of the coding activities and experience with students of various ages during prior pilot studies, we only included participants within the age range of 11 to 17. Participants were not paid to take part in the study, but could keep the agents they developed online on the ConvoBlocks website and use them later. Participants did not need prior programming experience or an Alexa-enabled device to participate. The only requirements were a computer with Zoom installed and access to the internet. The research study was approved by the researchers' Institutional Review Board prior to the study.

\begin{acks}
\end{acks}

\bibliographystyle{ACM-Reference-Format}
\bibliography{biblio}

\end{document}